\documentclass[preprint]{elsarticle}
\usepackage{amsfonts}
\usepackage{amssymb}
\usepackage{amsmath}
\usepackage{mathalpha}
\usepackage{mathtools}
\usepackage{mathrsfs}

\usepackage{tikz}
\usetikzlibrary{shapes,positioning,arrows}

\usepackage{graphicx}
\usepackage{xcolor}
\usepackage{amssymb}
\usepackage{amsthm}
\usepackage{amsmath}

\theoremstyle{plain}
\newtheorem{theo}{Theorem}[section]
\newtheorem{lem}[theo]{Lemma}
\newtheorem{proposition}[theo]{Proposition}
\newtheorem{cor}[theo]{Corollary}

\theoremstyle{definition}
\newtheorem{defn}{Definition}[section]

\newtheorem{ex}[theo]{Example}

\newtheorem{remark}[theo]{Remark}

\newcommand{\gen}[1]{{\langle #1 \rangle}}
\def\Z{{\mathbb Z}}

\usepackage{lineno}

\journal{Discrete Mathematics}

\begin{document}

\begin{frontmatter}

\title{On the structure of repeated-root polycyclic codes over local rings}

\author[1]{Maryam Bajalan }  
\ead{maryam.bajalan@math.bas.bg}
\author[2,v]{Edgar Mart\'inez-Moro}
\fntext[v]{Maryam Bajalan is supported by the Bulgarian Ministry of Education and Science, Scientific Programme ``Enhancing the Research Capacity in Mathematical Sciences (PIKOM)", No. DO1-67/05.05.2022.}
\ead{edgar.martinez@uva.es}
\fntext[v]{Second author was supported in part by Grant TED2021-130358B-I00 funded by MCIN/AEI/10.13039/501100011033 and by the “European Union NextGenerationEU/PRTR” }
\author[3]{Reza Sobhani}
\ead{r.sobhani@sci.ui.ac.ir}
\author[4]{Steve Szabo}
\ead{Steve.Szabo@eku.edu}
\author[5,w]{Gülsüm Gözde Yılmazgüç}
\ead{gozdeyilmazguc@trakya.edu.tr}
\fntext[w]{Last author is supported by TÜBİTAK within the scope of 2219 International Post Doctoral Research Fellowship Program with application number 1059B192101164. Her work was completed  while she visited the Institute of Mathematics of University of Valladolid (IMUVa). She thanks the IMUVa for their kind hospitality.}

\address[1]{Institute of Mathematics and Informatics, Bulgarian Academy of Science,  Acad. G. Bonchev Str. Bl. 8, 1113, Sofia, Bulgaria}
\address[2]{Institute of Mathematics, University of Valladolid, Castilla, Spain}
\address[3]{Department of Applied Mathematics and Computer Science, Faculty of Mathematics and Statistics, University of Isfahan,
Isfahan, Iran}
\address[4]{Department of Mathematics \& Statistics, Eastern Kentucky University, USA}
\address[5]{
Ipsala Vocational College, Trakya University, Edirne, Turkey}

\begin{abstract}
This paper provides  the Generalized Mattson Solomon polynomial  for repeated-root polycyclic codes over local rings that gives an explicit decomposition of them in terms of idempotents which completes the single root study in \cite{trans}.  It also states some  structural properties of repeated-root polycyclic codes over finite fields in terms of matrix product codes. Both approaches provide  a description of the $\perp_0$-dual code for a given polycyclic code.\end{abstract}

\begin{keyword}
Polycyclic code \sep Duality \sep Repeated-root codes \sep Mattson-Solomon transform \sep Matrix-product codes

 \MSC 94B15
\sep
13M10
 \sep  15B33
\end{keyword}

\end{frontmatter}

\section{Introduction} 
Polycyclic codes over a finite local ring $R$  were introduced in \cite{SteveCycli}  and they are described as  ideals on the quotient ring $R[x]/\langle f(x) \rangle$
with $f(x)\in R[x]$. These codes generalize the well-known classes of cyclic and constacyclic codes. Polycyclic codes over finite fields have been studied from several points of view, we will be especially interested in the so called $\perp_0$-duality (see  \cite{0dual,polyinv} and the references therein). Polycyclic codes  over chain rings have been studied in different directions, see for example  \cite{Fot20,Steve,Shi13,Shi17}.
In \cite{trans} the authors made a generalization where the ring is a finite commutative local   ring and  the polynomial defining the ambient space has simple roots.  That paper proposed a transform approach that generalizes the classical Mattson-Solomon (Fourier)  transform  in finite fields.

On the other side, several papers have been devoted to explain the matrix product code structure of repeated-root cyclic and constacyclic codes over finite fields, see for example \cite{Sobhani,MPCao}, and over some finite chain rings \cite{MP3}.

In this paper, we complete the study on the Mattson-Solomon   transform approach in \cite{trans} for polycyclic codes over finite local rings in the case that the defining polynomial has repeated-roots. We also give a matrix product code structure that describes repeated-root polycyclic codes over finite fields. In both cases, we provide expressions for the $\perp_0$-dual code of a given polycyclic code. 

The structure of the paper is as follows.  In  Section~\ref{sec:prev}, some preliminaries are given on finite commutative local rings, on  the Hasse derivative of a polynomial over a finite local ring  and on the  Generalized Discrete Fourier Transform. Section~\ref{GMSP} provides  the Generalized Mattson Solomon polynomial(GMS)  for  polycyclic codes over local rings that gives an explicit decomposition of them in terms of idempotents. In Section~\ref{MPC}, we state some  structural properties of repeated-root polycyclic codes over finite fields in terms of matrix product codes. In both  Section~\ref{GMSP} and Section~\ref{MPC}, we give a description of the $\perp_0$-dual code of a given polycyclic code.

\section{Preliminaries}\label{sec:prev}
Throughout the paper, $R$ will denote a finite local ring of characteristic $q=p^r$ for a prime $p$ and a positive integer $r$, $\mathfrak m$
will denote the maximal ideal of $R$ and $\mathbb F_q = R/\mathfrak m$ the finite residue field
of $R$. It is well-known that $R$ is trivially complete and thus Hensel, i.e. every element of $R$ is nilpotent or a unit and  $\mathfrak m$ is a nilpotent ideal.  We denote by $\bar\cdot$ the natural polynomial ring morphism $\bar\cdot : R \rightarrow (R/\mathfrak m)$ and, abusing notation, we will use it also for polynomial rings acting on the coefficients $\bar\cdot : R[x] \rightarrow (R/\mathfrak m)[x]=\mathbb{F}_q[x]$.
Let $\mathcal J$ denote the set of all polynomials $f$ in $R[x]$ such that
$\bar f$ has distinct zeros in the algebraic closure of $\mathbb{F}_q$,  a polynomial in $ \mathcal J$ has distinct zeros in local extensions of
$R$, $\mathcal R_f=R[x]/\langle f\rangle$
(where $f$ is monic) is a separable local extension ring
if and only if $f$ is an irreducible polynomial in $\mathcal J$, and the polynomials
in $\mathcal J$ admit a unique factorization into irreducible polynomials and a polynomial in $\mathcal J$ has no multiple roots in any
local extension of $R$. In the rest of the paper, unless other thing is stated, $f$ will denote a polynomial in  $\mathcal J$ and $F=f^m$ for a non-negative integer $m$ (in some sections $m=p^k$ where $p$ is the characteristic of $R$).

  \subsection{Hasse derivative and Generalized Discrete Fourier Transform}
  The Generalized Discrete Fourier Transform (GDFT) for repeated-root cyclic codes over a finite field $\mathbb F_q$ of characteristic $p$ ($p$ a prime)  of  length $N=np^k,$ where $(n,p)=1,$  was defined by Massey in \cite{massey}.  After that,  the definition is generalized for   quasi-cyclic and quasi-twisted codes over finite fields  in \cite{sole} and \cite{Jia}, respectively. 
  In those references, the Hasse derivative of polynomials  over finite fields plays an important role. For more information about the Hasse derivative of polynomials over fields we refer the reader to \cite{massey, Hir}.
  
 In this section, let  $R$  denote a commutative finite unitary ring and $p(x)=\sum_{i=0}^{n} p_ix^i\in R[x]$ be a polynomial. For $k\in\{0,1,\ldots, n\}$, the $k^{th}$ formal derivative of $p(x)$ is defined as  $p^{(k)}(x)=k!\sum_{i=0}^{n}{{i}\choose{k}}p_ix^{i-k}$,
  and the $k^{th}$ Hasse derivative of $p(x)$  is defined as  $p^{[k]}(x)=\frac{1}{k!}p^{(k)}(x)$  \cite[page $363$]{Hasse}, i.e.
    $$p^{[k]}(x)=\sum_{i=0}^{n}{{i}\choose{k}}p_ix^{i-k}=\sum_{i=0}^{n-k}{{i+k}\choose{k}}p_{i+k}x^{i}.$$ The following result holds directly from the definition and straightforward computations.
  
  \begin{lem} Let $p(x)$ and $q(x)$ be two polynomials in $R[x].$ 
  \begin{enumerate}
  \item $(p+q)^{[k]}(x)=p^{[k]}(x)+q^{[k]}(x).$
      \item (Taylor expansion)  If $p(x)$ is of degree $n$ and    $\lambda$ is an arbitrary element in $R,$ then
      $p(x+\lambda)=\sum_{k=0}^{n} p^{[k]}(\lambda)x^k.$
      \item (Product rule)  $(pq)^{[k]}(x)=\sum_{i=0}^{k}p^{[i]}(x)q^{[k-i]}(x)$.
  \end{enumerate}
  \end{lem}
 
From Now on, let  simple-root polynomial $f(x)=(x-\alpha_0)(x-\alpha_1)\ldots(x-\alpha_{n-1})\in \mathcal J$  has  $n$ fixed ordering distinct roots  $\alpha_0,\alpha_1,\ldots,\alpha_{n-1}$ in an extension ring $R'$ of $R.$  
 Recall that the Discrete Fourier Transform (DFT) of an $n$-tuple $(g_0,g_1,\ldots,g_{n-1})$ is $(g(\alpha_0),g(\alpha_1),\ldots,g(\alpha_{n-1})), $ where 
$g(x)=g_0+g_1x+\ldots +g_{n-1}x^{n-1}\in R[x]/\langle f(x)\rangle$  ; see \cite{trans}.
 \begin{defn}\label{N1}
  Let $F(x)=((x-\alpha_0)(x-\alpha_1)\ldots(x-\alpha_{n-1}))^m=(f(x))^m$ be a repeated-root monic polynomial  in  $R[x]$ of degree $N=nm$  and  $g(x)=\sum_{i=0}^{N-1}g_{i}x^{i}\in R[x]/\langle F(x)\rangle.$ We define the Generalized Discrete Fourier Transform (GDFT) of $g(x)$  as
 $$
 \begin{bmatrix}
 g(\alpha_0) & g(\alpha_1) & \ldots & g(\alpha_{n-1})\\
 g^{[1]}(\alpha_0) & g^{[1]}(\alpha_1) & \ldots & g^{[1]}(\alpha_{n-1})\\
 \vdots & \vdots &\ldots &\vdots\\
 g^{[m-1]}(\alpha_0) & g^{[m-1]}(\alpha_1) & \ldots & g^{[m-1]}(\alpha_{n-1})
 \end{bmatrix},
 $$
 where $g^{[i]}$ is the $i^{th}$-Hasse derivative for all $1\leqslant i\leqslant m-1.$
  \end{defn}

 \begin{ex}\label{N2}
 Suppose that $F(x)=x^6-3x^5+3x^4-x^3\in \mathbb Z_4[x],$ which is  decomposed  over $\mathbb Z_{16}$ as  $F(x)=(x-1)^3(x-12)^3.$ If $g(x)=1+2x^3+x^4+3x^5\in \mathbb Z_4[x]/\langle F(x)\rangle,$ then $g^{[1]}=2x^2+3x^4$ and $g^{[2]}=2x+2x^2+2x^3.$ Therefore, the GDFT of $n$-tuples related to $g(x)$ is
 $$
 \mathrm{GDFT}(g)=
 \begin{bmatrix}
 g(1) & g(12)\\ 
 g^{[1]}(1) & g^{[1]}(12)\\ 
 g^{[2]}(1) & g^{[2]}(12)
   \end{bmatrix}
 =
 \begin{bmatrix}
7 & 1\\ 
5 & 0\\ 
6 & 8
 \end{bmatrix}.
 $$ 
 
 \end{ex}
\subsection{Generalized Vandermonde matrices}

Let $\alpha_0,\alpha_1,\ldots,\alpha_{n-1}$ be a fixed ordering of the roots of polynomial  $f(x)=(x-\alpha_0)(x-\alpha_1)\ldots(x-\alpha_{n-1})\in R[x]$ in the extension ring $R'$ of $R.$

For $0\leqslant i\leqslant N-1,$ take  $p_i(x)=x^i$ and  construct the  $N\times m$ matrix 
   $$R(x)=
 \begin{bmatrix}
 p_0(x) & p_0^{[1]}(x) & \ldots & p_0^{[m-1]}(x)\\ 
  p_1(x) & p_1^{[1]}(x) & \ldots & p_1^{[m-1]}(x)\\ 
  \vdots & \vdots  &\ldots &\vdots\\
 p_{N-1}(x) & p_{N-1}^{[1]}(x) & \ldots & p_{N-1}^{[m-1]}(x)
  \end{bmatrix}.
 $$
  In fact, $ij$-entry of $R(x)$ is ${{i-1}\choose  {i-j}}x^{i-j}$ for $i\geqslant j$ and zero otherwise. The \textit{generalized Vandermonde matrix} related to the roots $\alpha_0,\alpha_1,\ldots,\alpha_n$ of the repeated-root polynomial $F(x)=(f(x))^m$ of degree $N=nm$ over  a local ring $R$ is defined by
    $$V=V(\alpha_0,\alpha_1,\ldots,\alpha_{n-1})=[R(\alpha_0)\,\,R(\alpha_1)\,\,\ldots R(\alpha_{n-1})].$$
\begin{ex}
 If $F(x)=(x-\alpha_0)^3(x-\alpha_1)^3$ then 
  $$
  V=[R(\alpha_0)\,\,R(\alpha_1)]=
  \begin{bmatrix}
1 & 0 & 0 & 1 & 0&0\\
\alpha_0 & 1 & 0 & \alpha_1 & 1&0\\
\alpha_0^2 & 2\alpha_0 & 1 & \alpha_1^2 & 2\alpha_1&1\\
\alpha_0^3 & 3\alpha_0^2 & 3\alpha_0 & \alpha_1^3 & 3\alpha_1^2& 3\alpha_1\\
\alpha_0^4 & 4\alpha_0^3 & 6\alpha_0^2 & \alpha_1^4 & 4\alpha_1^3& 6\alpha_1^2\\
\alpha_0^5 & 5\alpha_0^4 & 10\alpha_0^3 & \alpha_1^5 & 5\alpha_1^4& 10\alpha_1^3
\end{bmatrix}.
$$
\end{ex}
Note that  if $m=1,$ the  generalized Vandermonde matrix is compatible with the usual Vandermonde matrix related to $F(x).$
  The determinant of $V$ is $\prod_{0\leqslant i<j\leqslant n-1}(\alpha_i-\alpha_j)^{n_in_j};$ see \cite{GV}. Thus   $V$ is  invertible   in the local ring $R$ if and only if $\alpha_i-\alpha_j$ is a unit in $R$ if and only if $\overline{\alpha_i}\ne \overline{\alpha_j};$ see Lemma 2.5 in \cite{Norton}. Therefore  $V$ is invertible if and only if  $\overline{\alpha_i}\ne \overline{\alpha_j}$ for all $i\neq j.$  Note that since throughout the paper, it is assumed that
   $f\in\mathcal J,$ then $\bar{f}$ has distinct roots $\overline{\alpha_i}$ for $0\leqslant i \leqslant n-1.$ Thus $V$ will always be  an invertible matrix.
 
 Let $F(x)= x^N-\sum_{i=0}^{N-1} F_{i}x^{i}$ and  $C_F$ be the Companion matrix related to $F(x),$ i.e.
 $$
 C_F=
 \begin{bmatrix}
 0 & 1 & 0 & \ldots & 0\\ 
  0 & 0 & 1 & \ldots & 0\\ 
  \vdots & \vdots & \vdots &\ldots &\vdots\\
   0 & 0 & 0 & \ldots & 1\\ 
   F_0 &F_1 & F_2 & \ldots & F_{N-1}
 \end{bmatrix}.
 $$
 It is a well-known fact  that $F(x)$ is the characteristic polynomial of $C_F$,  since the polynomial  $F(x)$ has repeated-roots, the matrix $C_F$ is not  diagonalizable, but it can be reduced to a very simple form by means of the  generalized Vandermonde matrix. Let us
 denote the  Jordan form of the companion matrix  $C_F$ by $J_F,$ i.e.  a diagonal block matrix  with $n\times n$  blocks  so that each block  has  roots on the diagonal, $1$ on the superdiagonal and other entries are zero. If $V$ is invertible, then the Companion matrix is reduced to $C_F=VJ_FV^{-1}.$

\section{Generalized Mattson Solomon polynomial}\label{GMSP}

Let  $V$ be  the usual  Vandermonde matrix  related to the distinct elements $\alpha_0,\ldots,\alpha_{n-1}$ and $f(x)=\prod_{i=0}^{n-1} (x-\alpha_i)$. For a given  $g(x)=\sum_{i=0}^{n-1} g_{i}x^{i}$ in $R[x]/\langle f(x)\rangle,$ the Mattson Solomon polynomial of $g(x)$ is  
  \begin{equation}\label{10}
    \mathrm{MS}(g)=\sum_{i=0}^{n-1} g(\alpha_{i})x^{i} =[g_0\,g_1\,\ldots g_{n-1}]V[1\,x\,\ldots\,x^{n-1}]^T.   
  \end{equation} Note that the map $\mathrm{MS}$  is well defined in the quotient space $R[x]/\langle f(x)\rangle$ (see \cite{trans} for a complete account on it).
  Now, 
  let $F(x)=f(x)^m$ be a repeated-root polynomial of degree $N=mn$
over the local ring $R$   and  fix an  ordering on  distinct roots $\alpha_0,\ldots,\alpha_{n-1}.$
Let us consider the quotient polynomial ring $\mathcal R=\left (\frac{{R'}[y]}{\langle y^m\rangle},\cdot\right)$, where $\cdot$ is the ordinary polynomial multiplication modulo $y^m$.
\begin{theo}\label{m3}
The map 
$$\begin{array}{lccl}
    \mathrm{MS}: &  \left(\frac{{R}[x]}{\langle F(x)\rangle},\bullet\right ) & \longrightarrow &\left(\frac{{\mathcal R}[x]}{\langle f(x) \rangle},\star\right )\\[1em]
   &g(x) & \mapsto  &
\sum_{j=0}^{n-1}\left(\sum_{i=0}^{m-1}  g^{[i]}(\alpha_j)y^i\right) x^j
\end{array}$$
is a ring injective homomorphism, where $\bullet$ denotes ordinary polynomial multiplication
modulo $F(x)$ and $\star$ denotes the component-wise multiplication modulo $f(x)$.
\end{theo}
\begin{proof} {First, we will show that the mapping is well-defined. Given two representatives $h(x), g(x)$ of an element in $\frac{{R}[x]}{\langle F(x)\rangle}$, that is $g(x)-h(x)=k(x) f(x)^m$, for $0\leq i \leq m-1$. We have by applying the product rule that
$$
g^{[i]}(x)-h^{[i]}(x)=\sum_{j=0}^i k^{[i]}(x) (f(x)^m)^{[i-j]} .
$$
But $(f(x)^m)^{[i-j]}=(i-j)!(f(x)^m)^{(i-j)}$ (the usual derivative of $f(x)^m$) which is indeed $0$ mod $f$ for $0\leq i \leq m-1$. Therefore for $0\leq i\leq m-1$ one has that $g^{[i]}(x)$, $h^{[i]}(x)$ provide the same values when evaluated at $\alpha_j$, $j=0,\ldots, n-1$.}

Let  $g(x)=\sum_{i=0}^{N-1} g_{i}x^{i}\in \frac{{R}[x]}{\langle F(x)\rangle} $ and $V$ be the  generalized Vandermonde matrix related to roots $\alpha_0,\ldots,\alpha_{n-1}$. Consider the column vector
$$u=\left[1\,y\,\ldots\,y^{m-1}\,\,\,x\,\,\,\,xy\,\ldots\,xy^{m-1}\,\ldots\,x^{n-1}\,\,\,\,x^{n-1}y\,\ldots\,x^{n-1}y^{m-1}\right]^{\mathrm{tr}},$$ where $\mathrm{tr}$ denotes the transpose of the vector. Then we have that $\mathrm{MS}(g)$ is given by
\begin{equation*} 
\begin{split}
& \left[ {g({\alpha _0})\,\,\,{g^{[1]}}({\alpha _0})\ldots {g^{[m - 1]}}({\alpha _0})\ldots g({\alpha _{n-1}})\,\,\,{g^{[1]}}({\alpha _{n-1}})\, \,\,\ldots {g^{[m - 1]}}({\alpha _{n-1}})} \right]u\\
 & = \left [g_0\,g_1\,\ldots g_{N-1}\right]Vu.
\end{split}
\end{equation*}
Since  the matrix  $V$ is invertible, then $\mathrm{MS}$ is injective.
 Now it is enough  to show that $\mathrm{MS}(g\bullet h)=\mathrm{MS}(g)\star \mathrm{MS}(h)$ that follows applying the product rule of the Hasse derivative, we can easily check that 
   $\mathrm{MS}(g)\star \mathrm{MS}(h)$ can be computed as 
 $$\sum_{i=0}^{n-1} \left( \left( \sum_{j=0}^{m-1}g^{[j]}(\alpha_i)y^{j}\right)\cdot\left( \sum_{j=0}^{m-1}h^{[j]}(\alpha_i)y^{j}\right)\right) x^i = \sum_{i=0}^{n-1} \left(\sum_{j=0}^{m-1}(gh)^{[j]}(\alpha_i) y^j \right) x^i.$$
\end{proof}
Note that  the mapping in the above theorem gives the ordinary Mattson-Solomon transform when applied to a simple-root polynomial. Thus abusing the notation,
 we will denote both the same. We will call the map $\mathrm{MS}$ in the above theorem  \emph{the Generalized Mattson Solomon} map associated to $F$.
 
\begin{ex}( Example \ref{N2} Cont.) Let $m=3, \, n=2, \, f(x)=x^2-x\in\mathbb Z_4[x]$ and   $\mathcal R=\mathbb Z_{16}[y]/\langle y^3\rangle. $ Then
$$\mathrm{MS}(g(x))=(7+5y+6y^2)+(1+8y^2)x\in \mathcal R[x]/\langle f(x)\rangle.$$
\end{ex}
\begin{remark}
Theorem \ref{m3} states that every repeated-root polycyclic code
is isomorphic  to an ideal  in a  bivariable polynomial ring, since $
\frac{{\mathcal R}[x]}{\langle f(x) \rangle} \cong \frac{R'[x,y]}{\langle f(x),y^m\rangle}.$  
\end{remark}
\begin{lem}\label{MB17}
 The  map $\mathrm{MS}$ in Theorem~\ref{m3} is equivalent to  each of the following mappings.
 \begin{equation}\label{MB43}
    \begin{array}{lccl}
    \mathrm{MS}: &  \left(\frac{{R}[x]}{\langle F(x)\rangle}\,,\bullet\right ) & \longrightarrow &\left(\frac{{\mathcal R}[x]}{\langle f(x) \rangle}\,,\star\right )\\[1em]
   &g(x) & \mapsto  & \sum_{i=0}^{n-1} g({\alpha _i} + y) x^i
\end{array}
 \end{equation}
 and 
 \begin{equation}\label{MB44}
 \begin{array}{lccl}
    \mathrm{MS}: &  \left(\frac{{R}[x]}{\langle F(x)\rangle}\,,\bullet\right ) & \longrightarrow &\left( {\frac{{{\frac{R'[u]}{\langle (u-1)^m\rangle}} [x]}}{{\langle f(x)\rangle }}\,, \star } \right)\\[1em]
   &g(x) & \mapsto  & \sum_{i=0}^{n-1} g(u{\alpha _i}) x^i.
\end{array}
\end{equation}
 \end{lem}
\begin{proof}

     Since $y^m=0,$ by  the Taylor expansion for the Hasse derivative,  we get 
 $g(\alpha_i+y)=\sum_{j=0}^{m-1}g^{[j]}(\alpha_i)y^j.$ Thus
 $\mathrm{MS}(g)=\sum_{i=0}^{n-1} g({\alpha _i} + y) x^i$, which gives the mapping \eqref{MB43}. The set $\{\alpha_0(y+1)-y,\ldots , \alpha_{n-1}(y+1)-y\}$ are roots of $F(x),$   since $y^m=0.$  Put $u=y+1$ and use the mapping \eqref{MB43} to find $\mathrm{MS}(g(x)).$ Now,   $R'[u-1]\cong R'[u]$ provides  the mapping \eqref{MB44}.

\end{proof}
\begin{remark}
     Note that in the case  $F(x)$ is a simple-root polynomial (i.e  $m=1$), we get $y=0$ and $u=1$. Hence, the two mappings  presented in the previous lemma
   are  compatible with the Mattson Solomon mapping given in \cite{trans}. 
    \end{remark}
    \begin{remark} In definition  \ref{N1}, we define the GDFT  for polycyclic codes of length $N=mn$ over rings as a generalization of  the GDFT for repeated-root cyclic codes   of  length $N=np^k$  over fields presented by  Massey in \cite{massey}. Now we are able to present other definitions of the GDFT  associated with the mappings in Lemma \ref{MB17}: 
 \begin{equation}\label{MB45}
\begin{array}{cccc}
\mathrm{GDFT}:& R^N &\longrightarrow& \mathcal{R}^n\\
&(g_0,g_1,\ldots,g_N) &\mapsto& (g(\alpha_0+y),\, g(\alpha_1+y),\, \ldots , g(\alpha_{n-1}+y))
\end{array}
   \end{equation} 
and 
 \begin{equation}\label{MB46}
\begin{array}{cccc}
\mathrm{GDFT}:& R^N &\longrightarrow& \mathcal A^n\\
&(g_0,g_1,\ldots,g_N) &\mapsto& (g(u\alpha_0),\, g(u\alpha_1),\, \ldots , g(u\alpha_{n-1})),
\end{array}
   \end{equation} 
   where $\mathcal A=\frac{R'[u]}{\langle (u-1)^m\rangle}.$ Note that \eqref{MB45} is compatible with the definition of the DFT given in \cite{trans}. 
    \end{remark}
\section{The decomposition of the ambient space}
We will start by studying the ring   $\mathcal R$ defined in the previous section.
\begin{lem}[Corollary 3.8 in \cite{Mc}]\label{m1}
Let $R$ be a local ring and $g\in R[x]$ be a monic irreducible polynomial. Then $R[x]/\langle g(x)^n\rangle$ is a local ring for any positive integer $n$.
\end{lem}
\begin{lem}\label{m2}
Let $S$ be a Galois extension of the local ring $R$. Then 
\begin{enumerate}
    \item $S$ is the unramified local ring, i.e. $R$ and $S$ has the same maximal ideal. 
    \item If $f\in R[x]$ is square-free, then $f$ has distinct zeros in the local extension $S$. 
    \item $S$ is an $R$-free module generated by roots of $f$.
\end{enumerate}
\end{lem}
\begin{proof}
See Theorems 3.15, 3.18, 5.11 in \cite{Mc}
\end{proof}

\begin{cor}\label{m4}
Let $R'$ be the  Galois extension of  the local ring $R$  containing $n$ distinct roots of the polynomial $f(x)=\prod_{i=0}^{n-1}(x-\alpha_i).$  Then $\mathcal R=R'[y]/\langle y^m\rangle$ {is a local ring.}
\end{cor}
The proof of the corollary follows from the fact that $R$ is local, $R'$ is a Galois extension and applying Lemmas~\ref{m1}~and~\ref{m2}. Then, from the counting argument in \cite{local}, if we count  the elements in $\mathcal R$ that are $p^{m s}$, and   the number of zero divisors in $\mathcal R$ is $p^{c+ (s-1) m}$ where $p^c$ is the number of zero divisors of $R$, therefore from \cite[Theorem 1]{local} $\mathcal R$, is a local ring. Furthermore, note that   $\mathcal R$ is   also a chain ring if and only if $R$ is a finite field. This follows from the fact that the maximal ideal of $\mathcal R$ is   $\langle m,y\rangle$ where $m$ is the generator of the maximal ideal of $R$ and it is principal if $p$ is the characteristic of $R$.  

\subsection{Decomposition of the codes}\label{ss:decomp}
 
In this section, we are going to find a decomposition of the ambient space $\frac{{ R}[x]}{\langle F(x) \rangle}$. %
Recall that the ring  $\left(\frac{{\mathcal R}[x]}{\langle f(x) \rangle}\,, \star\right)$ is  equipped with the component-wise product  and the ring  $\left(\frac{{ R}[x]}{\langle F(x) \rangle}, \bullet\right)$ is equipped  with the  ordinary polynomial product. Let us denote  $\frac{{ R}[x]}{\langle F(x) \rangle}$ by $R_F.$  Let $f=f_1f_2\ldots f_r,$ where $f_1,f_2\ldots f_r$  are distinct monic irreducible polynomials.  We will  define a relation on the set of indices $I=\{0,1,\ldots,n-1\}$ as follows: 
$i\sim j$  if and only if $\alpha_i,\alpha_j$ are roots of the same polynomial $f_k,$ i.e $f_k(\alpha_i)=f_k(\alpha_j)=0$. Therefore $I$ will be partitioned into the disjoint classes  $I_k$ related to $f_k$.  

 From now on in Subsection~\ref{ss:decomp}, we will consider the MS-map in Theorem~\ref{m3} extended to $R^\prime$ 
 $$
    \mathrm{MS}:   \left(R^\prime_F=\frac{{R^\prime}[x]}{\langle F(x)\rangle},\bullet\right )  \longrightarrow \left(\frac{{\mathcal R}[x]}{\langle f(x) \rangle},\star\right )$$or, what is the same, consider a polynomial $f(x)$ which completely splits in linear factors in the ring we are working on.
    
It is easy to see that, again for cardinality reasons,  it is now an isomorphism and 
we  can define $E_i=\mathrm{MS}^{-1}\left( \sum_{j\in I_i}  x^{j}\right)$.  The pre-images $\{E_1,\ldots, E_r\}$ will provide us the primitive idempotents, more precisely:
\begin{proposition}\label{m7}\
\begin{enumerate}
    \item Each $E_i$ is a primitive idempotent.
\item $E_iE_j=0$  for $i\neq j,$ and $\sum_{i=1}^{r}E_i=1$
\item The only idempotents in $R_F$ are in the form $\sum_{j\in S}E_j$ for some $S\subseteq\{1,2,\ldots, r\}.$
\item  $R^\prime_F \cong \oplus_{i=1}^r \langle {{E_i}}\rangle  \cong  \oplus_{i=1}^r \frac{R_F}{{\left\langle {1 - {E_i}} \right\rangle }}$
\end{enumerate}
    \end{proposition}
\begin{proof}\
\begin{enumerate}
    \item Note that $x^j\star x^j=x^j$ for all $0\leq j\leq n-1$, thus   $E_i^2=\mathrm{MS}^{-1}\left( \sum_{j\in I_i}  x^{j}\right)=E_i.$
    To check that $E_i$ is primitive, let $E_i=A(x)+B(x),$ where $A(x)$ and $B(x)$ are primitive idempotents in $R_F.$ Denote $\mathrm{MS}(A(x))= \sum_{k=0}^{n-1} a_k  x^{k}=a(x)$ and $\mathrm{MS}(B(x))=\sum_{k=0}^{n-1} b_k  x^{k}=b(x)$. Then $\sum_{j\in I_i}  x^{j}=a(x)+b(x)=\sum_{i=0}^{n-1}(a_i+ b_i)  x^{i},$ and hence  $a_k+b_k=0$ for $k\notin I_i$ and $a_k+b_k=1$ otherwise.  Since $A(x), B(x)$ are idempotent elements,  $a(x), b(x)$ are also idempotent elements  in $\mathcal R[x]$.  According to   componenet-wise  multiplication  in $\mathcal R[x],$ we conclude that $a_i$ and $b_i$ are idempotent elements in $\mathcal R$ for all $0\leqslant i\leqslant n-1.$ Now since $R_y$ is local, $a_k,b_k\in\{0,1\}$ for all $0\leq k\leq n-1$. So if we let $a(x)$ and $b(x)$ be  equal to the unit element $\sum_{j=0}^{n-1}  x^{j}$ in $ R_y[x],$ then  $a_k=b_k=1$ for all $0\leqslant k \leqslant n-1,$ which is a contradiction with $a_k+b_k=1$ for  $k\in I_i.$ Therefore, $a(x)$ and $b(x)$ are not the unit element in $ R_y[x].$  On the other hand, we have    $a(x)=0$ or $b(x)=0,$ that is $A(x)=0$ or $B(x)=0$.
        \item  For $i\neq j,$  $I_i$ and $ I_j$ are disjoint  and hence  $E_iE_j=0.$   Moreover, since   $\sum_{i=0}^{n-1}  x^{i}$  is the  unit element  of  $\mathcal R[x]$ we get
        $$1=\mathrm{MS}^{-1}(\sum_{i=0}^{n-1} x^{i}).$$
    \item  Clearly, to obtain the idempotents in $R_F^\prime,$ it is necessary to study idempotents in $MS(R_F^\prime)=\frac{\mathcal R[x]}{\langle f(x)\rangle}.$ Let $a(x)=\sum_{k=0}^{n-1}a_kx^k$ be an idempotent element in $\frac{R_y[x]}{\langle f(x)\rangle}.$ We get $\sum_{k=0}^{n-1}a_kx^k=a(x)=a(x)^2=\sum_{k=0}^{n-1}a_k^2x^k.$ Thus $a_k=a_k^2$ for all $0\leqslant k\leqslant n-1$ and since $R_y$ is local, we have $a_k\in\{0,1\}$ for all $0\leqslant k\leqslant n-1$. If we let   $S=\{i\mid a_i\neq 0\},$ then $a(x)=\sum_{i\in S} x^i$ and 
    $A(x)=\mathrm{MS}^{-1}(a(x))=\sum  _{i\in S}E_i.$
    \item The first isomorphism  follows from the fact that $\{E_1,\ldots,E_r\}$ is  the  set of pairwise primitive orthogonal idempotents.
      To prove the second isomorphism we  define $\theta: R_F\to \langle E_i\rangle $ via $g\mapsto gE_i.$ Let $gE_i=0.$ Then $g=g(1-E_i)+gE_i=g(1-E_i),$ and hence $\ker \theta =\langle 1-E_i\rangle,$ which gives the result. 
\end{enumerate}
\end{proof}
This provides the following description of the codes in terms of the idempotents in the case of a ring of prime characteristic.
\begin{proposition}\label{MB14}
Let $R^\prime$ be a  local ring with prime characteristic $p$ and  $N=np^k.$  Then 
\begin{enumerate}
    \item If $f_i(x)=\prod_{j\in I_i}(x-\alpha_{j})$  then  $(f_i(x))^{p^k}=1-E_i$.
    \item If the ideal $C$ of $R^\prime_F$ has an idempotent generator, then $C$ is generated by $\prod_{i\in S}(f_i(x))^{p^k}$ for some $S\subseteq \{1,2,\ldots,r\}.$
\end{enumerate}
\end{proposition}
\begin{proof}\
\begin{enumerate}
    \item  $1-E_i=\mathrm{MS}^{-1}(\sum_{i=0}^{n-1}x^{i})-\mathrm{MS}^{-1}(\sum_{i=0}^{n-1}d_{i}x^{i})= \mathrm{MS}^{-1}(\sum_{i=0}^{n-1}e_{i}x^{i})$ such that $e_i\notin I_i.$  On the other hand, recall that $\alpha_i-\alpha_j$ is a unit in $R^\prime$ if and only if $\overline{\alpha_i}\ne \overline{\alpha_j}.$  Since 
   $f\in\mathcal J,$  $\bar{f}$ has distinct roots $\overline{\alpha_i}$ for $0\leqslant i \leqslant n-1,$  and  we get

    \begin{align*}
               (f_i(\alpha_j+y))^{p^k}&=
               (\alpha_j+y-\alpha_{i_1})^{p^k}\ldots  (\alpha_j+y-\alpha_{i_t})^{p^k}\\
                              &=((\alpha_j-\alpha_{i_1})^{p^k}+y^{p^k})\ldots  ((\alpha_j-\alpha_{i_t})^{p^k}+y^{p^k})\\
                               &=(\alpha_j-\alpha_{i_1})^{p^k}\ldots  (\alpha_j-\alpha_{i_t})^{p^k}\\
                                          &=    \begin{dcases}
        0 & j\in I_i, \\
       \hbox{unit} & j\notin I_i. \\
           \end{dcases}
    \end{align*}
    Thus 
    $\mathrm{MS}((f_i(x))^{p^k})=\sum_{j=0}^{n-1}(f_i(\alpha_0+y))^{p^k}x^j\in\langle \sum_{j\notin I_i}x^j\rangle=\mathrm{MS}(1-E_i).$
  Now  since $\mathrm{MS}$ is injective, the result holds.
   \item The only idempotent elements in $R^\prime_F$  are in the form $\sum_{i\in K} E_i$ for some subset $K$ of $\{1,2,\ldots,r\}.$ By the fact that $E_i$'s are orthogonal we have
   $$\sum_{i\in K} E_i=1-\sum_{i\notin K} E_i=\prod_{i\notin K}(1-E_i)=\prod_{i\notin K}(f_i(x))^{p^k}.$$
    Now it is enough to take $S=K^c.$

\end{enumerate}
\end{proof}

\begin{cor}\label{MB15}
Let $R^\prime$ be a  local ring with prime characteristic $p$ and  $N=np^k$ where $f$ completelly splits.  Then 
$$\frac{R^\prime[x]}{\langle F(x)\rangle}=\frac{R^\prime[x]}{\langle (f(x))^{p^k}\rangle}\cong\bigoplus\limits_{i=1}^{r}\frac{R^\prime[x]}{\langle (f_i(x))^{p^k}\rangle}$$
\end{cor}
\begin{proof}
Part (4) of Proposition \ref{m7}, part (1) of Proposition \ref{MB14} and  the Third Isomorphism Theorem  give the proof.
\end{proof}

\begin{remark} Note that in this section (Section~\ref{ss:decomp}) we have considered codes over the ring $R^\prime_F$, if we want to restrict ourselver to $R_F$ we must consider subring subcodes that behave as subfield subcodes in the field case, for a reference on they, their Galois closure and a Delsarte's like theorem in the chain ring case see~\cite{Trace}.  \end{remark}

\subsection{$\perp_0$  duality}\label{ssec:duality}
 Consider the following inner product over the ring $R_F=\frac{{ R}[x]}{\langle F(x) \rangle}$
\begin{equation}
    \langle g_1(x), g_2(x) \rangle_{(0)}=  g_1g_2(0),\quad  g_1(x), g_2(x) \in  R_F. 
\end{equation} 
We will denote the dual of the  polycyclic code $C\subseteq R_F$ associated with this inner product by $C^{\perp_0}$ given by 
\begin{equation*}
C^{\perp_0}=\{g(x)\in R_F \mid  \langle g(x), h(x) \rangle_{(0)}=0, \hbox{for all}\,\, h(x)\in C\}.
\end{equation*}
\begin{theo}\label{MB50}
Let $C$ be a polycyclic code of length $N=np^k$ in $R_F.$ If $F_0$ is an  invertible element in $ R$, then 
\begin{enumerate}
    \item\label{MB40} The inner product $\langle \, , \rangle_{(0)}$ is non-degenerate.
    \item \label{MB41} $C^{\perp_0}=\mathrm{Ann}(C)$, where $\mathrm{Ann}$ stands for the annhilator ideal. 
    \item $C^{\perp_0}$ is a polycyclic code. 
\end{enumerate}

\end{theo}
\begin{proof}\
\begin{enumerate}
    \item We must show that the orthogonal of $R_F$ is zero. Let $g=g_0+g_1x+\ldots+g_{N-1}x^{N-1}\in R_F$ and $ \langle g, x^i \rangle_{(0)}=0$ for all $0\leqslant i\leqslant N-1.$ From  $ \langle g, 1 \rangle_{(0)}=0$ we conclude  $g_0=0.$ Also,  by considering $0= \langle g, x^i \rangle_{(0)}=g_{N-i}F_0$ for all $1\leqslant i\leqslant N-1$  and invertibility $F_0$ we obtain $g_{N-i}=0$, i.e. $g=0$.  

\item { Let $h(x)\in  \mathrm{Ann}(C)$, therefore $h(x)g(x)= 0$ for all $g(x)\in C$ and hence $hg(0)=0$, i.e. $h(x)\in C^{\perp_0}$.  Thus $\mathrm{Ann}(C)\subseteq C^{\perp_0}.$} Conversely, let $h\in C^{\perp_0}$ and  $g\in C$ be an  arbitrary element. Hypothesis $ 0=\langle g, h \rangle_{(0)}=hg(0)$  implies that $x^ihg(0)=0$ for all $0\leqslant i\leqslant N-1.$ Now by part \eqref{MB40} we have $hg=0,$ which gives the result.
\item  It is obvious by part \eqref{MB41}.
\end{enumerate}
\end{proof}
\begin{remark}\label{rem:unit}
In the literature of simple-root polycyclic codes over $R[x]/\langle f(x)\rangle$, it is always  assume that $f_0$ is a unit in the ring $R,$ see 
\cite{SteveCycli,Fot20}. Because this assumption is guaranteed  that every left polycyclic code is right polycyclic and as a result we get ride of studying left and right at the same time.   In this paper, we always  assume that $F(0)=F_0$, the constant term of the polynomial $F$,  is a unit in $R$. Because this assumption is guaranteed  that the dual of every polycyclic code ($C^{\perp_0}$) is again polycyclic (also in our previous paper in simple-root case \cite{trans} we have assumed that $f_0$ is a unit in order to have a polycyclic dual). 
\end{remark}

We now define another inner product over  $R_F:$
\begin{equation}
    \langle g_1(x), g_2(x) \rangle_{\mathrm{MS}}=  \mathrm{MS}(g_1)\star \mathrm{MS}(g_2),\quad  g_1(x), g_2(x) \in  R_F,
\end{equation} 
As usual we will denote the dual of the  polycyclic code $C\subseteq R_F$ associated with this inner product by $C^{\perp_{\mathrm{MS}}},$ which  is naturally defined as 
\begin{equation}
    C^{\perp_{\mathrm{MS}}}=\{g\in R_F\mid \mathrm{MS}(g)\star \mathrm{MS}(c)=0\,\hbox{ for all} \,c\in C\}.
 \end{equation}
 The following result shows how one can check the annhilator duality in terms of the Mattson Solomon transform.
    \begin{theo}
    For the polycyclic code $C$ over $R_F,$ we have $\mathrm{Ann}(\mathcal C)= \mathcal C^{\perp_{\mathrm{MS}}}.$
    \end{theo}
    \begin{proof}
    Since the Mattson-Solomon mapping is an  injective morphism  we have 
 \begin{equation*}
     gc=0\iff \mathrm{MS}(gc)=0 \iff \mathrm{MS}(g)\star \mathrm{MS}(c)=0
      \end{equation*}
      which implies $\mathrm{Ann}(\mathcal C)= \mathcal C^{\perp_{\mathrm{MS}}}.$
    \end{proof}

 \begin{remark}
 Note that all the results in Subsection~\ref{ssec:duality} are given in the ring $R_F$ since only injectivity of the $\mathrm{MS}$ map is needed, so we do not need to consider the ring $R^\prime_F$. 
 \end{remark}

\subsection{A note on multivariable codes}
In \cite{trans}, the  Mattson Solomon map  for several variable serial codes over chain rings presented. That construction was based on the decomposition of the tensor product of  the two $R$-modules $R[x_1]/\langle f_1(x_1)\rangle $ and $R[x_2]/\langle f_2(x_2)\rangle $ in terms   of the tensor product of powers of  their related companion matrices $E_f$ and $E_g$ and their simultaneous diagonalization by the matrix $V_{f_1}\otimes V_{f_2}$ where $V_{f_i}$ is  Vandermonde matrix corresponding to $f_i, \, i=1,2$.  In the principal ideal case, one of the defining polynomials is a repeated-root one, say $f_1(x)=f(x_1)^m,$ and the remainder ones should be non-repeated-root polynomials and $R$ is a Galois ring, see  \cite{Edgar}. In that later is the case, we can provide a Mattson Solomon transform in terms of the Generalized Vandermonde matrices in the same fashion as in \cite{trans}.

Multivariable   codes over the ring $R$ are ideals of the quotient ring   $\mathscr{R}=R[x_1,\ldots x_w]/\langle  t_1(x_1),\ldots,t_w(x_w)\rangle.$ If all polynomials  $t_1(x_1),\ldots t_w(x_w)$ are simple-roots, then these codes are  called serial multivariate codes, and otherwise they are called modular multivariate codes. The transform approach to the serial case   over local rings was studied in \cite{trans}. Note that serial multivariate  codes are well-behaved because  they can be regarded as principal ideals in $\mathscr{R}.$ This property is not generally true in the modular case. In the case $r>2,$  $\mathscr{R}$ is principal ideal ring if and only if $R$ is a Galois ring and the number of polynomials for which $\bar{t_i}(x_i)$ is not square-free is at most one, see  \cite[Theorem 1]{Edgar}.

For the sake of simplicity, all results in this section will be proved for $w = 2$ and can be straight forward worked out for $w > 2$. Let $R$ be a Galois ring,  $f(x_1)$ a polynomial of degree $n$ over $R$ with  distinct simple-roots $\alpha_0,\ldots,\alpha_{n-1}$ in an extension ring $R'_1,$ and  $F(x_1)=(f(x_1))^m$ a polynomial  of degree $N=nm$.  Moreover, let $g(x_2)$ be a polynomial of degree $M$ over $R$ with distinct  simple-roots $\beta_0,\ldots,\beta_{M-1}$ in an extension ring $R'_2.$ Let $V$ be the generalized Vandermonde matrix related to $\alpha_0,\ldots,\alpha_{n-1}$ and $v$ be the usual Vandemonde matrix related to $\beta_0,\ldots,\beta_{M-1}$. Consider the tensor product
$$ v\otimes V= \begin{bmatrix}
V & \ldots & V \\
\beta_0 V & \ldots & \beta_{M-1}V \\
\vdots & \ldots & \vdots \\
\beta_0^{M-1} & \ldots & \beta_{M-1}^{M-1}V 
\end{bmatrix}.  $$
Since $det(v\otimes V)=det(v)^{M}det(V)^N$ and $v, V$ are invertible,   then $v\otimes V$ is invertible. 
A polynomial $p(x_1,x_2)\in R[x_1,x_2]/\langle F(x_1), g(x_2) \rangle $ can de written as $p(x_1,x_2)=\sum_{j=0}^{M-1}p_j(x_1)x_2^j,$ where $p_j(x_1)=\sum_{i=0}^{N-1}p_{i,j}x_1^i.$ Relate the vector 

\begin{equation*}
   p=(p_{0,0},\, p_{1,0},\,\ldots,\, p_{N-1,0},\, p_{0,1},\, p_{1,1},\,\ldots,\, p_{N-1,1},\,\ldots,\, p_{0,M-1},\, p_{1,M-1},\,\ldots,\, p_{N-1,M-1}) 
\end{equation*} to the polynomial $p(x_1,x_2).$ It can be easily seen  that the product of the vector $p$ and  matrix $v\otimes V$ is as follows:
\begin{multline*}
p(v\otimes V)=\big (p(\alpha_0+y, \beta_0),\, \ldots,\, p(\alpha_{n-1}+y, \beta_0),\, p(\alpha_0+y, \beta_1),\, \ldots,\, p(\alpha_{n-1}+y, \beta_1),\,\\ \ldots,\, p(\alpha_0+y, \beta_{M-1}),\, \ldots,\, p(\alpha_{n-1}+y, \beta_{M-1}) \big). 
\end{multline*}
Take $R''=R'_1 + R'_2.$ Clearly,  $p({\alpha _i} + y, \beta_j)\in R''$ for all $0\leqslant i\leqslant n-1$ and $0\leqslant j\leqslant M-1.$  Define the  multivariable Mattson-Solomon transform for
modular multivariable  codes as
 \begin{equation*}
    \begin{array}{lccl}
    \mathrm{MS}: &  \left(\frac{{R}[x_1,x_2]}{\langle F(x_1), g(x_2)\rangle}\,,\bullet\right ) & \longrightarrow &\left(\frac{ R''[x_1, x_2]}{\langle f(x_1), g(x_2) \rangle}\,,\star\right )\\[1em]
   &p(x_1, x_2) & \mapsto  & \sum\limits_{i=0}^{n-1}\sum\limits_{j=0}^{M-1} p({\alpha _i} + y, \beta_j) x_1^{i}x_2^j
\end{array}
 \end{equation*}
 where  $\bullet$ denotes ordinary polynomial multiplication
modulo $F(x_1),\, g(x_2)$ and $\star$ denotes the component-wise multiplication modulo $f(x_1),\,g(x_2)$. Obviously, the mapping  $\mathrm{MS}$ is a ring homomorphism and since $v\otimes V$ is invertible,  $\mathrm{MS}$ is also injective.


\section{Matrix-Product Structure of Certain  Polycyclic Codes}\label{MPC}

We prove the structure of some repeated-root polycyclic codes with the help of matrix-product codes in the paper \cite{Sobhani}. From now on, we will consider repeated-root polynomials just over the finite field $\mathbb F_q$, where $q=p^r$ where $p$ is a prime number.   Let $f(x)\in \mathbb F_{p^r}[x]$ be a simple-root polynomial of degree $n$ and of order $e$, i.e.
$e$ is the smallest integer for which $f(x) | x^e -1$ and $\mathrm{gcd}(p, e) = 1$. Let
$f (x) =\prod_{i=1}^{s}f_ i (x)$ be the unique factorization of $f(x)$ into distinct irreducible polynomials over $\mathbb F_{p^r}[x]$. Then, we  have  $f (x^{p^k}) =\prod_{i=1}^{s} f_ i (x^{p^k})$ 
and  for each $1 \leq i \leq s$, there exists an irreducible polynomial
$g_i (x)$ in $\mathbb F_{p^r}[x]$ such 
that $f_i (x^{p^k} ) = g_i (x)^{p^k}$. From now on, we will assume that $R$ is the ring 
\begin{equation}
  R=\mathbb F_{p^r}[x]/\langle f(x^{p^k})\rangle =  \mathbb F_{p^r}[x]/\left\langle \left(\prod_{i=1}^{s} g_i(x)\right)^{p^k}\right\rangle
\end{equation}
 and we will have that  $N = np^k$.
One can write any element $a(x)\in R$ as $a_0 (x) + a_1(x) x^{p^k} +\ldots + a_{n-1}(x) x^{(n-1)p^k}$, where $a_i(x)\in {\mathbb{F}}_{p^r}[x]$.
Let $S$ be the ring $\mathbb F_{p^r}[x,y]/\langle x^{p^k} -y, f (y)\rangle$. We
have the following straight forward results.

\begin{lem}\label{L1}  Any ideal  of the ring $R$ is principally generated by a divisor of $f(x^{p^k} )$.
In fact, it is of the form   $\langle G(x)\rangle$, where $G(x)=\prod_{j=1}^s g_i(x)^{i_j}$
and $0 \leq i_j \leq p^k$.
\end{lem}
\begin{remark}
Note that in the case of cyclic codes, the above ideals  give us the so-called monomial like codes in \cite{HFSE}.
\end{remark}

\begin{lem}\label{L2}
The map $\varphi:R \rightarrow S$ given by
$\varphi\left(\sum_{i=0}^{n-1} a_i(x)x^{i p^k}\right)=a(x,y)= \sum_{i=0}^{n-1} a_i(x)y^i$
is a ring isomorphism.
\end{lem}
\noindent Now we will consider the ring
\begin{equation}
T=\mathbb F_{p^r}[x,y]/\langle{x^{p^k}-1,f(y)}\rangle=\left(\mathbb F_{p^r}[x]/\langle{x^{p^k}}-1\rangle\right)[y]/\langle f(y)\rangle,
\end{equation}
and  denote as $W$ the ring $W=\mathbb F_{p^r}[x]/\langle{x^{p^k}-1\rangle}$. Note that  $W$  is a finite chain ring whose maximal ideal is $\langle (x-1)\rangle$.

\begin{lem}\label{L3}
The map
$\psi:S\rightarrow T$ defined by $\psi(a(x,y))=a(y^{e'}x,y)$ is a ring isomorphism, where $e'$ is the inverse of $p^k$ in $\Z_e$.
\end{lem}
As an easy corollary we have the following.
{\begin{cor}\label{C1}
 The code  $C$ is a   polycyclic code  in $\mathbb F_{p^r}[x]/\langle f(x^{p^k})\rangle $  if and only if
$\mu(C)=\psi(\varphi(C))$ is a  polycyclic code in  $W[y]/\langle f(y)\rangle$. 
\end{cor}

Therefore, since $W$ is a chain ring we can apply  \cite[Theorem 3.5]{Sergio}  and we get the following  unique $( x - 1)$-adic expansion of the code $C$ (Note that we have also a description of a system of generators of a polycyclic code over a chain ring  in  \cite[Theorem 4.4]{Salagean2} and its generalization in \cite[Theorem 3.13]{EdgarRua}).

\begin{proposition}\label{P1}
Any polycyclic code $C$ in $W[y]/\langle{f(y)}\rangle$  is of the form
\begin{eqnarray*}
C & = & \langle{h_0(y),(x-1)h_1(y),\dots,(x-1)^{p^k-1}h_{p^k-1}(y)}\rangle,
\end{eqnarray*}
where $h_{p^k-1}(y)\mid h_{p^k-2}(y)\mid \dots\mid h_0(y)\mid f(y)$ over $\mathbb F_{p^r}$.
Moreover,  we have $$C=\bigoplus_{i=0}^{p^k-1}(x-1)^iC_i,$$ where for $0\le i\le p^k-1$, $C_i=\langle{h_i(y)}\rangle$
is a polycyclic code   in $\mathbb F_{p^r}[y]/\langle f(y)\rangle$ and $C_0\subseteq C_1\subseteq \cdots\subseteq C_{p^k-1}$.
\end{proposition}

Note that the ideal defining $C$ over the ring $W$ is a single generated and the generator can be derived from the polynomials $h_i(x)$ in the above expression (see the proof of \cite[Theorem 3.13]{EdgarRua}).The following theorem follows directly
\begin{theo}\rm\label{T1}
Let $C=\langle{g_1(x)^{i_1}g_2(x)^{i_2}\cdots g_r(x)^{i_r}}\rangle$. Then we have 
$$\mu(C)=\bigoplus_{i=0}^{p^k-1}(x-1)^iC_i$$
where $C_i$ is a simple-root polycyclic code with respect to $f(y)$ over $\mathbb F_{p^r}$. In fact we have $C_i=\gen{k_i(y)}$, where $k_i(y)=\prod_{j\in A_i}g_j(y)$ and $A_i=\{1\le j\le r\ |\ i_j>i\}$.
\end{theo}

The following definition introduces matrix product codes in this work. Matrix-product codes over some classes of rings have been studied in several works, see for example \cite{MP1,MP2,MP3,MP4}, but they did not consider the $\perp_0$-orthogonality.

\begin{defn}\label{matrixproduct}
 Let $A = [a_{ij}]$ be an $\alpha \times  \beta$ matrix
with entries in $\mathbb F_{p^r}$
and  let $C_1,\ldots C_{\alpha}$ be codes of length $n$ over $\mathbb F_{p^r}.$ The matrix-product code $[C_1, \ldots , C_{\alpha}]\cdot A$ is the set of
all matrix products $[c_1,\ldots, c_{\alpha}]A$, where $c_i\in C_i$, defined by

   \begin{align}\label{MB12}
   [c_1,\ldots,c_{\alpha}]\cdot A &= [c_1,\ldots,c_{\alpha}]\begin{bmatrix}
a_{11} & a_{12} & \ldots & a_{1\beta} \\
a_{21} & a_{22} & \ldots & a_{2\beta} \\
\vdots & \vdots & \ldots & \vdots \\
a_{\alpha 1} & a_{\alpha 2} & \ldots & a_{\alpha \beta} 
\end{bmatrix}  \\
&= [a_{11}c_1 + a_{21}c_2 + \ldots + a_{\alpha 1}c_{\alpha}, a_{12}c_1 + a_{22}c_2 + \ldots + a_{\alpha 2}c_{\alpha},\nonumber\\
& \qquad\qquad\ldots  , a_{1\beta}c_1 + a_{2\beta}c_{2} + \ldots + a_{\alpha \beta}c_{\alpha}]. \nonumber
\end{align} 
\end{defn}

\begin{lem}[Proposition 2.9 \cite{Blackmore}]\label{MB20}
 If a matrix consisting of some $\alpha$ columns of $A$ is non-singular and $C = [C_1,\ldots, C_{\alpha}]\cdot A,$ then $\mid C\mid=\mid C_1\mid\ldots \mid C_{\alpha}\mid.$
\end{lem}

\begin{defn}[Definitions~1 and ~2  in \cite{Sobhani}] \label{CYC} $\quad $

\begin{itemize} 
    \item $J$ to be the $p^k \times p^k$ matrix whose $(i,p^k-i+1)$-th entry $(1\leq i \leq p^k)$ is equal to $1$ and other entries are equal to zero, $P$ to be the $p^k \times p^k$ matrix whose  $(i,j)$-th entry $(1\leq i,j \leq p^k)$ is equal to $\binom{i-1}{j-1}$ mod $p$, $Q$ to be the $p^k \times p^k$ matrix whose  $(i,j)$-th entry is equal to $(-1)^{(i+j)}\binom{i-1}{j-1}$ mod $p$ and $\mathrm{CYC}(p,k)$ to be $JQJ$.
    \item For $0\le i\le N-1$ we will write $i=ap^k+j$ where $0\le a\le n-1$, $0\le
j\le p^k-1$. We define the permutation $\sigma$ on
$\{0,1,\dots,N-1\}$ as $\sigma(i)=jn+a$. 
\end{itemize}
\end{defn}

\begin{lem} \label{non-sing}  $A=\mathrm{CYC}(p,k)$ is a non-singular matrix.
\end{lem}
\begin{proof}
Matrix $\mathrm{CYC}(p,1)$ is upper triangular with exactly $p-i$ zeros in the  column $i$ and ones in the diagonal,   and  $A=\mathrm{CYC}(p,k)=\bigotimes_{i=1}^k \mathrm{CYC}(p,1)$ by \cite{Sobhani}. Since  the tensor product of two upper triangular matrices is again upper triangular the result follows. 
\end{proof}

\begin{theo}\label{T2}
 Let $C$ be a polycyclic code  in $\mathbb F_{p^r}[x]/\langle (f(x))^{p^k}\rangle$  and  $\mu(C)=\bigoplus_{i=0}^{p^k-1}(x-1)^iC_i$, then we have that
 $$\sigma(C)=[C_{p^k-1},C_{p^k-2},\dots,C_0]\cdot \mathrm{CYC}(p,k).$$
\end{theo}
\begin{proof} Assume $a(x)=\sum_{i=0}^{n-1}a_i(x)x^{i\cdot p^k}\in C$, then
$\varphi(a(x))=\sum_{i=0}^{n-1}a_i(x)y^{i}$ and
hence
$\psi(\varphi(a(x)))= \sum_{i=0}^{n-1} a_i(y^{e^\prime}x)y^{i}.$
If
$\sigma(a(x))=b(x)=\sum_{i=0}^{p^k-1} x^ib_i(x^{p^k})$
then we have
$\psi(\varphi(a(x)))= \sum_{i=0}^{p^k-1} (y^{e'}x) ^ib_i(y)$

On the other hand, we can write
\begin{eqnarray*}
b_0(y)+(y^{e'})xb_1(y)+\cdots+y^{(p^k-1)e'}x^{(p^k-1)}b_{p^k-1}(y) =\\
b_0(y)+(y^{e'})(x-1+1)b_1(y)+\cdots+y^{(p^k-1)e'}(x-1+1)^{(p^k-1)}b_{p^k-1}(y) =\\
\sum_{i=0}^{p^k-1} \left(\sum_{j=0}^{i}{i\choose j}(x-1)^j\right) y^{je'}b_i(y)=\\
\sum_{j=0}^{p^k-1}y^{je'}\left(\sum_{i=j}^{p^k-1}{i\choose j}b_i(y)\right)(x-1)^j.
\end{eqnarray*}
For  $0\le j\le p^k-1$, let us denote by  $c'_j(y)=y^{je'}\sum_{i=j}^{p^k-1}{i\choose j}b_i(y)$, and $c_j(y):=\sum_{i=j}^{p^k-1}{i\choose j}b_i(y)$. Hence
$\sum_{j=0}^{p^k-1}c'_j(y)(x-1)^j\in\psi(\varphi(C))$ and since
$\psi(\varphi(C))=\bigoplus_{i=0}^{p^k-1}(x-1)^iC_i,$
we have  $c'_j(y)\in C_j$. { But $C_j$ is a polycyclic code and $y$ is a unit element because we assume that $f_0 $ is a unit (see Remark~\ref{rem:unit}). Hence $c_j(y)\in C_j$ as well.} Now we have
$$[c_0(y),c_1(y),\dots,c_{p^k-1}(y)]=[b_0(y),b_1(y),\dots,b_{p^k-1}(y)]\cdot
P,$$ where  $P$ is an invertible matrix whose inverse is
the matrix $Q$. Therefore we have
$$[c_0(y),c_1(y),\dots,c_{p^k-1}(y)]\cdot Q=[b_0(y),b_1(y),\dots,b_{p^k-1}(y)],$$
and it follows $\sigma(C)\subseteq [C_0,C_1,\dots,C_{p^k-1}]\cdot
Q$ and since both of the sets have the same size, we have
$\sigma(C)=[C_0,C_1,\dots,C_{p^k-1}]\cdot Q$.
Using similar arguments as those used in \cite{Sobhani}, we get
$
\sigma(C) = [C_{p^k-1},C_{p^k-2},\dots,C_0]\cdot \mathrm{ CYC}(p,k),
$
and the proof is now completed.
\end{proof}

\begin{remark} Note that
if we consider $C$ as a cyclic code of length $np^k$ over the field $\mathbb F_{p^m}$ in \cite{Sobhani}, a permutation $\pi$ is provided such that
$$\pi(C)=[C_{p^k-1},C_{p^k-2},\dots,C_0]\cdot \mathrm{CYC}(p,k).$$
It can be easily checked that, in general, $\pi\neq \sigma$, where $\sigma$ is the permutation defined above, while the codes $C_i$, $0\leq i\leq p^k-1$, are the same.  Therefore we have two permutations for which $\pi(C)=\sigma(C)$ or equivalently $\pi^{-1}\circ\sigma\in\mathrm{Aut}(C)$, the group of automorphism of the code $C$. The reason for getting different permutation in this case is related to the different kinds of isomorphisms we have considered. In fact, in~\cite{Sobhani} the mapping considered was
$$\frac{\mathbb F_{p^m}[x]}{\langle x^{np^k}-1\rangle} \stackrel{\sim}{\longrightarrow} \frac{F_{p^m}[x,y]}{\langle x^{n}-y, y^{p^k}-1\rangle},
$$ while in this paper we have considered the isomorphism $$\frac{\mathbb F_{p^m}[x]}{\langle x^{np^k}-1\rangle}\stackrel{\sim}{\longrightarrow} \frac{F_{p^m}[x,y]}{\langle x^{p^k}-y, y^{n}-1\rangle}.
$$ 
\end{remark}

Since the matrix $\mathrm{CYC}(p,1)$ is a Non-Singular by Columns matrix (NSC matrix) (see \cite{Sobhani} for a definition), Proposition~2 in \cite{Asch}  implies the following corollary involving the minimum Hamming  distance $d_i$ of each of the component codes $C_i$  and the distance of the code $d(C)$.

\begin{cor}\label{C2}
Let $C$ be a polycyclic code  in $\mathbb F_{p^r}[x]/\langle (f(x))^{p^k}\rangle$ such that $\mu(C)=\bigoplus_{i=0}^{p^k-1}(x-1)^iC_i$,
then we have
$$d(C)={\rm min}\{
p^ k d_{p^k -1},(p^k -1)d_{p^k-2},\dots,d_0\},$$ where
$d_t=d(C_t)$ and $t=0,1,...,p^k-1$. 
\end{cor}

\subsection{Duality}
The annihilator dual of a matrix-product code  can be also  explicitly described in terms
of matrix-product codes. First we will introduce the following auxiliary result.

\begin{lem}\label{MB23}
The isomorphism $\mu$ introduced in Corollary~\ref{C1} is a $\perp_0$-duality preserving map, i.e, $\mu(C^{\perp_0})=(\mu(C))^{\perp_0}$.
\end{lem}
\begin{proof}
For all $p(x)$ and $q(x)$ in $\mathbb F_{p^r}[x]/\langle f(x^{p^k})\rangle,$ it is easy to see that
\begin{equation}\label{MB232}
    \langle p(x),q(x)\rangle_0=0\iff \langle \mu (p(x)),\mu (q(x))\rangle_0=0.
\end{equation}
Let $p(x)\in C^{\perp_0}.$ By  Equation \eqref{MB232}, we have $\langle \mu(p(x)),\mu(q(x))\rangle_0=0$ for all $q(x)\in C,$ i.e $\mu(p(x))\in (\mu(C))^{\perp_0},$ which gives $\mu(C^{\perp_0})\subseteq (\mu(C))^{\perp_0}$. Conversely, let $z\in (\mu(C))^{\perp_0}$. Then $\langle z,\mu(p(x))\rangle_0=0$ for all $p(x)\in C.$ Using Equation \eqref{MB232}, we get $\langle \mu^{-1}(z),p(x)\rangle_0=0$ for all $p(x)\in C,$ which implies $\mu^{-1}(z)\in C^{\perp_0},$ i.e, $z\in \mu(C^{\perp_0})$.
\end{proof}


We will need the following Theorem to prove some results relating the $\perp_0$-dual of the matrix product code in terms of the  of the $\perp_0$-duals of their constituent codes. For a matrix $A$ we will  denote its transpose as $A^\mathrm{tr}$.

\begin{theo}\label{MB21}
Let $D_0,\ldots , D_{p^k-1}$ be polycyclic codes over $\mathbb F_{p^r}[x]/\langle f(x^{p^k})\rangle.$ Then 
$$\big([D_{p^k-1},\ldots ,D_1,D_0]\cdot A\big)^{\perp_0}=[D^{\perp_0}_{p^k-1},\ldots, D^{\perp_0}_1,D^{\perp_0}_0]\cdot(A^{-1})^{\mathrm{tr}}.$$
\end{theo}
\begin{proof}
We claim that 
\begin{equation}\label{MB18}
[\textrm{Ann}(D_{p^k-1}),\ldots, \textrm{Ann}(D_0)]\cdot({A^{-1}})^{\textrm{tr}}\subseteq \textrm{Ann} \big([D_{p^k-1},\ldots,D_0]\cdot A\big).
\end{equation}
Indeed, let $z=[z_{p^k-1},\ldots, z_0]\cdot ({A^{-1}})^{\textrm{tr}}\in [\textrm{Ann}(D_{p^k-1}),\ldots, \textrm{Ann}(D_0)]\cdot({A^{-1}})^{\textrm{tr}}. $ Note that $z$ is a row vector. If we consider the product of two row vector $v,w$ as $v.w=vw^{\textrm{tr}},$ then   for an arbitrary element $x=[x_{p^k-1},\ldots, x_0]\cdot A\in [D_{p^k-1},\ldots,D_0]\cdot A$ we have 
\begin{align*}
  z.x&=\big([z_{p^k-1},\ldots, z_0]\cdot({A^{-1}})^{\textrm{tr}}\big)\cdot\big(A^{\textrm{tr}}\cdot [x_{p^k-1},\ldots,x_0]^{\textrm{tr}}\big) \\
  &=[z_{p^k-1},\ldots, z_0]\cdot[x_{p^k-1},\ldots,x_0]^{\textrm{tr}}=0.
  \end{align*}
  Using the above claim, we get 
$$[(D_{p^k-1})^{\perp_0},\ldots, (D_0)^{\perp_0}]\cdot ({A^{-1}})^{\textrm{tr}}\subseteq  \big([D_{p^k-1},\ldots,D_0]\cdot A\big)^{\perp_0}.$$
By Lemmas~\ref{MB20},~\ref{non-sing} it follows 
\begin{align*}
    \mid [(D_{p^k-1})^{\perp_0},\ldots, (D_0)^{\perp_0}]\cdot({A^{-1}})^{\textrm{tr}}\mid&=
    \mid (D_{p^k-1})^{\perp_0} \mid\ldots\mid (D_0)^{\perp_0}\mid  \\
    &= \frac{\mid \mathbb F_{p^r}\mid^n}{\mid D_{p^k-1}\mid}\ldots \frac{\mid \mathbb F_{p^r}\mid^n}{\mid D_{0}\mid}\\
    &=\frac{\mid \mathbb F_{p^r}\mid^{p^kn}}{\mid D_{p^k-1}\mid\ldots \mid D_{0}\mid}\\
    &=\frac{\mid \mathbb F_{p^r}\mid^{p^kn}}{\mid [D_{p^k-1},\ldots,D_0]\cdot A\mid }\\
    &=\mid \big([D_{p^k-1},\ldots,D_0]\cdot A\big)^{\perp_0}\mid, 
\end{align*}
which gives the proof.
\end{proof}

\begin{cor}\label{MB22}
$$\big([D_{p^k-1},\ldots ,D_1,D_0]\cdot \mathrm{CYC}(p,k)\big)^{\perp_0}=[D^{\perp_0}_{0},\ldots, D^{\perp_0}_{p^k-2},D^{\perp_0}_{{p^k-1}}]\cdot \mathrm{CYC}(p,k).$$
\end{cor}
\begin{proof}
\begin{align*}
\big([D_{p^k-1},\ldots, D_1,D_0]\cdot \mathrm{CYC}(p,k)\big)^{\perp_0}
&=[D^{\perp_0}_{p^k-1},\ldots, D^{\perp_0}_1,D^{\perp_0}_0]\cdot((\mathrm{CYC}(p,k))^{-1})^{\textrm{tr}}\\
&=[D^{\perp_0}_{p^k-1},\ldots, D^{\perp_0}_1,D^{\perp_0}_0]\cdot Q\\
&=[D^{\perp_0}_{0},\ldots, D^{\perp_0}_{p^k-2},D^{\perp_0}_{{p^k-1}}]\cdot JQ\\
&=[D^{\perp_0}_{0},\ldots, D^{\perp_0}_{p^k-2},D^{\perp_0}_{{p^k-1}}]\cdot \mathrm{CYC}(p,k).
\end{align*}
\end{proof}


Now, combining Theorem \ref{MB21} and Corollary \ref{MB22} we get the following result

\begin{cor} Let $C$ be a polycyclic code  in $\mathbb F_{p^r}[x]/\langle f(x^{p^k})\rangle$of such that 
 $\sigma(C)=[C_{p^k-1},C_{p^k-2},\dots,C_0]\cdot \mathrm{CYC}(p,k)$ as in Theorem~\ref{T1}.  Then 
$$\sigma(C^{\perp_0})=[C^{\perp_0}_{0},\ldots, C^{\perp_0}_{p^k-2},C^{\perp_0}_{{p^k-1}}]\cdot \mathrm{CYC}(p,k).$$
\end{cor}

\bibliographystyle{plain}
\bibliography{references.bib}
 \end{document}